\documentclass[twoside]{article}

\usepackage{microtype}
\usepackage{graphicx}
\usepackage{subcaption}
\usepackage{booktabs} 
\usepackage{url}

\usepackage{amsmath}
\usepackage{amssymb}
\usepackage{amsthm}
\usepackage{mathtools}
\usepackage{caption}
\usepackage{multirow}

\newtheorem{theorem}{Theorem}[section]
\newtheorem*{theorem*}{Theorem}

\newtheorem*{lemma*}{Lemma}
\newtheorem*{definition*}{Definition}
\newtheorem{assumption}{Assumption}
\newtheorem{prop}{Proposition}[section]
\newtheorem{corollary}[theorem]{Corollary}

\theoremstyle{definition}
\newtheorem{remark}{Remark}[section]
\newtheorem{setting}{Problem Setting}

\theoremstyle{definition}

\newcommand{\indep}{{\perp\!\!\!\perp}}

\usepackage[accepted]{aistats2022}
\usepackage[round]{natbib}

\begin{document}

%
\runningtitle{Outcome Assumptions and Balancing Weights}

%

\twocolumn[

\aistatstitle{Outcome Assumptions and Duality Theory for Balancing Weights}

\aistatsauthor{ David Bruns-Smith \And Avi Feller  }

\aistatsaddress{ UC Berkeley \And  UC Berkeley } ]

\begin{abstract}
We study balancing weight estimators, which reweight outcomes from a source population to estimate missing outcomes in a target population. These estimators minimize the worst-case error by making an assumption about the outcome model. In this paper, we show that this outcome assumption has two immediate implications. First, we can replace the minimax optimization problem for balancing weights
with a simple convex loss over the assumed outcome function class. Second, we can replace the commonly-made overlap assumption with a more appropriate quantitative measure, the minimum worst-case bias. Finally, we show conditions under which the weights remain robust when our assumptions on the outcomes are wrong.
\end{abstract}


    
    
    




\section{Introduction}

Using covariates to transfer outcome information from one setting to another is a central task in domain adaptation, observational causal inference, and missing data imputation.
These tasks share a common structure: we observe covariates and outcomes for a source data set and want to predict outcomes given covariates in a target data set, which might have a different covariate distribution than the source.
One standard approach is to reweight the source distribution to have a similar covariate distribution to the target. When the source and target distributions have common support, using the density ratio for weights leads to unbiased estimation, known as \emph{importance weighting} for domain adaptation under covariate shift \citep{sugiyama2007covariate} and \emph{inverse propensity score weighting} (IPW) for observational causal inference \citep{rosenbaum1983central}. 

Importance weighting has several drawbacks.
First, 
using the density ratio for weights can lead to extremely large variance and unstable estimation \citep{kang2007demystifying, cortes2010learning}. 
Second, the density ratio is notoriously difficult to estimate, and simple plug-in estimates do not guarantee covariate balance between the reweighted source distribution and target distributions \citep[see][]{eli2021balancingact}. 

Due to these drawbacks, in practice we would like to use weights with smaller variance than the density ratio that directly target a specified level of covariate balance \citep[e.g.,][]{gretton2009covariate, imai2014covariate}. In general, such a bias-variance trade-off only exists if we assume restrictions on the outcome model; without restrictions, only the density ratio can guarantee finite bias. This motivates the so-called \emph{minimax balancing weights} estimators, which we study in this paper. These estimators find the minimum dispersion weights that constrain the worst-case bias between groups over an outcome function class \citep[see][]{zubizarreta2015stable, hirshberg2019minimax, zhao2019covariate, kallus2020generalized}.



\subsection{Summary of the Paper}

We begin by reviewing existing balancing weights estimators, which achieve a smaller mean squared error than importance weighting by introducing an assumption on the outcome model. We argue that the outcome assumption implies two new results. 

First, we use convex duality to show that the minimax optimization problem for balancing weights can be replaced with a simple convex loss over the outcome function class. Our dual formulation in Section \ref{dualsection} shows that the minimax weights are always a (rescaled and recentered) function from that class. For example, if the outcomes are bounded, the corresponding weights will be bounded. If the outcome function belongs to an RKHS with some kernel, the corresponding weights will belong to an RKHS with the same kernel. The outcome assumption pins down the shape of the balancing weights. 


Second, we show that after making an outcome assumption, we do not need the density ratio to exist, i.e., we do not need to make the additional ``overlap'' assumption that is common in causal inference. Instead, there is an explicit quantity, the minimum achievable bias, which depends on the outcome function class, and which acts as a quantitative measure of the degree of overlap violations. We show that this measure can be more appropriate than an overlap assumption in finite samples for quantifying the underlying difficulty of the reweighting problem. 


Finally, given the central role of restrictions on the outcome model in both of the previous results, we briefly consider the setting in which this assumption is incorrect. In particular, we provide simple moment conditions under which we can retain a finite bound on the error when the true outcome model is not in the assumed class.

\subsection{Related work}




\textbf{Estimators that target balance.} 
Many reweighting estimators in causal inference explicitly target the discrepancy between source and target distribution, also known as \emph{balance} 
\citep{hainmueller2012entropy, zubizarreta2015stable, athey2018approximate, hirshberg2019minimax, zhao2019covariate, tan2020regularized, hazlett2020kernel, arbour2021permutation}. See \cite{eli2021balancingact} for a summary. 
The literature on domain adaptation also uses worst-case discrepancy between distributions \citep{mansour2009domain,gretton2009covariate, yu2012analysis, courty2014domain}. Some approaches learn representations that minimize these discrepancies \citep{ganin2016domain, shen2018wasserstein, pmlr-v130-assaad21a}. Closely related are estimators that 
target
the density ratio between two groups through a surrogate loss \citep{sugiyama2007direct, nguyen2010estimating, sugiyama2012density}.

\textbf{Overlap in causal inference and adversarial training.}
Many existing theoretical treatments of balancing weights require the density ratio to exist (called \emph{overlap} in causal inference), which is typically used for proving asymptotic consistency \citep[see][]{hirshberg2019minimax, kallus2020generalized}.
This assumption, however, can be highly restrictive especially in high-dimensions, as illustrated by \citet{d2021overlap}. 
See also \citet{khan2010irregular} for a discussion of the implications of overlap violations for causal inference.
The same topic arises in adversarial training. See, for example, \cite{dupuis2019formulation, birrell2020f}, who generalize $\phi$-divergences to distributions that do not have common support. This idea is applied to GANs in \cite{song2020bridging, glaser2021kale}.

\textbf{Domain adaptation and causal inference.} We emphasize that domain adaptation and causal inference are both special cases of the same problem setup. Related work combines ideas from these two literatures. For example, \cite{shalit2017estimating, johansson2020generalization} use integral probability metrics to estimate causal effects without the need for an overlap assumption. The same idea is used in \cite{kallus2020generalized} for matching estimators in causal inference. Other work has made the connection between causal inference and adversarial training \citep{yoon2018ganite, ozery2018adversarial}.  


\section{Problem Setup}

Let $X \in \mathcal{X}$ denote covariates, and $Y \in \mathbb{R}$ denote outcomes. We study the general class of problems with source and target populations, $P$ and $Q$, with different joint distributions over $X$ and $Y$. We observe $X$ in both populations, but only observe the outcomes, $Y$, for the source population, $P$. The goal is to estimate the missing mean in the target population, $\mathbb{E}_Q[Y]$. Many important problems share this structure, including causal inference and domain adaptation.

\begin{setting}[Causal Inference] 
Consider the causal inference setting with a binary treatment status variable, $T$, and potential outcomes $Y(0)$ and $Y(1)$. For the control group, we observe covariates $X$ and the potential outcome $Y(0)$. In the treated group, we still observe $X$ but do not observe $Y(0)$. Therefore, finding the average treatment effect on the treated is equivalent to solving the problem setup described above, where $P$ corresponds to the $T = 0$ population, $Q$ corresponds to the $T = 1$ population, and $Y$ corresponds to $Y(0)$.
\end{setting}

\begin{setting}[Domain Adaptation] 
Consider a classification task with features $X$, labels $Z$, and loss function $\ell$. In a source environment where we observe both $X$ and $Z$, we train a model $h$ for predicting $Z$ given $X$. We would like to estimate the average risk of our classifier in a new environment where we observe $X$ but not $Z$. This problem is equivalent to the setup above, where $P$ corresponds to the source environment, $Q$ corresponds to the target environment, and $Y$ corresponds to the loss, $\ell(h(X),Z)$.
\end{setting}

\subsection{Ignorability and Overlap}

To estimate the mean of $Y$ in $Q$ using the outcomes from $P$, we require some kind of regularity between the source and target populations. A common assumption is the \emph{ignorability} assumption (also called the \emph{covariate shift} assumption, or \emph{selection on observables}), which requires the relationship between covariates and outcomes to be the same across the two groups: 
\begin{assumption}[Ignorability]
\label{ignorability}
For all $x \in \mathcal{X}$,
\begin{align*} P(Y | X=x) = Q(Y | X=x). 
\end{align*}
\end{assumption}
Notice that in the causal inference setting described above, Assumption \ref{ignorability} is equivalent to the standard conditional independence assumption, $Y(0) \indep T | X$. 

Typically, Assumption \ref{ignorability} is paired with a requirement that the density ratio $dQ/dP$ exists, also known as \emph{overlap} or \emph{continuity} in different literatures: 
\begin{definition*}[Overlap]
We say that \emph{overlap} holds if $Q$ is absolutely continuous with respect to $P$.
\end{definition*}

\subsection{Importance Weighting}

In the special case where Assumption \ref{ignorability} and overlap hold, we can estimate the mean of the missing outcomes by reweighting the observed outcomes with the density ratio. This estimator is called \emph{importance weighting} or \emph{inverse probability weighting} (IPW) and is unbiased:
\begin{align*}
    \mathbb{E}_P \left[ \frac{dQ}{dP}(X) \hspace{0.1cm} Y \right] &= \mathbb{E}_P \left[ \frac{dQ}{dP}(X)\hspace{0.1cm}\mathbb{E}_P[Y|X] \right]\\
    &= \mathbb{E}_Q[ \mathbb{E}_P[Y|X]] = \mathbb{E}_Q[Y],
\end{align*}
where we use ignorability for the last equality. 

Importance weighting has two main drawbacks. First, the overlap assumption is very strong, especially in high dimensions \citep{d2021overlap}. But even if overlap holds in the super-population, in finite samples there are usually so-called practical overlap violations: regions of the covariate space that are well-represented in the target population, but very rare in the source population, leading to large importance weights.

\subsection{Mean Squared Error}

Large weights lead to large mean squared error. Consider arbitrary weights $w(X)$. We will expand the mean squared error (MSE) of $\mathbb{E}_P[w(X)Y]$ for estimating $\mathbb{E}_Q[Y]$ using the standard bias-variance decomposition. Define the \emph{outcome function}, $f_0(x) \coloneqq \mathbb{E}_P[Y|X=x] = \mathbb{E}_Q[Y|X=x]$ and likewise, let $\sigma^2_0(x)$ be the conditional variance of $Y$. 
Then,
\begin{align}
    \text{MSE}(w) &= \mathbb{E}_P[ (w(X) Y - \mathbb{E}_Q[Y])^2] \nonumber\\
    &= (\mathbb{E}_P[ w(X) Y ] - \mathbb{E}_Q[Y])^2 +  \text{Var}_P[ w(X) Y ] \nonumber\\
    &= (\mathbb{E}_P[ w(X) f_0(X) ] - \mathbb{E}_Q[f_0(X)])^2 \label{bias_term} \\
    &\;\;\; +  \mathbb{E}_P[w(X)^2 \sigma^2_0(X)]\label{var_term}  .
\end{align}
The MSE depends on two quantities: (1) the imbalance of the mean of the outcome function $f_0$ between the re-weighted source distribution and the target distribution; and (2) the variability of the weights under the source distribution, which amplifies the noise in the outcomes. With practical overlap violations in high dimensional problems, $w = dQ/dP$ can be enormous, and (\ref{var_term}) will result in a large MSE \citep[see, for example,][]{kang2007demystifying}. Balancing weights, introduced in Section \ref{balweights}, explicitly target the trade-off between bias and variance, as discussed extensively in \cite{kallus2020generalized}.

\subsection{Notation}
We now introduce formal notation used for the remainder of the paper. Let $(\mathcal{X}, \mathcal{S})$ be a measurable space.\footnote{To side-step topological issues, we assume that $\mathcal{X}$ is a separable Banach space.} Let $P$ and $Q$ be given probability measures on $(\mathcal{X}, \mathcal{S})$. Let $f_0$ be a real-valued measurable function on $\mathcal{X}$. Denote $\mathcal{M}(\mathcal{X})$ the space of signed finite measures on $(\mathcal{X}, \mathcal{S})$ and $\mathcal{M}(P)$ those absolutely continuous with respect to $P$. Denote $\mathcal{P}(\mathcal{X})$ the space of probability measures on $(\mathcal{X}, \mathcal{S})$ and $\mathcal{P}(P)$ those absolutely continuous with respect to $P$. 

With a slight abuse of notation, for measurable $f : \mathcal{X} \rightarrow \mathbb{R}$ and \emph{both} $M \in \mathcal{M}(\mathcal{X})$ and $M \in \mathcal{P}(\mathcal{X})$, we will write $\mathbb{E}_M[f] \coloneqq \int_\mathcal{X} f(x) dM(x)$. We assume $\mathbb{E}_P[|f_0|] < \infty$ and $\mathbb{E}_Q[|f_0|] < \infty$.



While our setting is quite general, it may be helpful for the reader to keep in mind the case where $\mathcal{X}$ is finite and discrete with cardinality $n$. In this case, $P$ and $Q$ are probability vectors of length $n$ and measurable functions are simply vectors in $\mathbb{R}^n$. Likewise, $\mathcal{M}(\mathcal{X})$ is just $\mathbb{R}^n$.

\section{Balancing Weights}\label{balweights}

In this section, we briefly review the existing work on balancing weights estimators and then introduce our main contributions. Balancing weights estimators find weights $w(X)$ with minimum dispersion, subject to a balance constraint between the target covariate distribution and the reweighted source distribution. In general, we will consider weights such that $\mathbb{E}_P[w(X)] = 1$, i.e., we always end up with the same ``size'' population that we started with. The problem of choosing weights can be reformulated as finding a measure $R \in \mathcal{M}(P)$ such that $\int_\mathcal{X} dR(x) = 1$, with $w \coloneqq dR/dP$. This corresponds to the intuition behind reweighting as creating a ``pseudo-population'' based on $P$ intended to match $Q$. We will therefore often use $w$ and $R$ interchangably. 

A simple balancing weights estimator might constrain the mean of the covariates to match within tolerance $\delta$, similarly to \cite{zubizarreta2015stable}. For example, let $\mathcal{X} = \mathbb{R}^d$. We could find the minimum variance $w$ such that
\begin{align} \Vert \mathbb{E}_R[ X ] - \mathbb{E}_Q[X] \Vert_2 \leq \delta\label{linearbal}, \end{align}
where, as a reminder, $w = dR/dP$. If the outcome function $f_0(X)$ is linear with bounded coefficients, then this constraint will bound the bias term (\ref{var_term}) and the tuning parameter $\delta$ lets us trade-off bias and variance to achieve a smaller MSE than the importance weights.

\subsection{Assumptions on the Outcome Function}\label{outcomeassumption}

More generally, we may not want to assume that $f_0$ is linear. But this presents a difficulty: without any further restrictions, for any $w \neq dQ/dP$, there always exists an adversarial $f_0$ that can make the bias term (\ref{bias_term}) arbitrarily large. Only the density ratio guarantees bounded bias for \emph{any} $f_0$, but often at the cost of high variance.

Therefore, in practical settings, we instead restrict the outcome function in order to control the error. To make progress, we  assume that $f_0$ belongs to some function class $\mathcal{F}$:
\begin{assumption}
\label{outcomes}
The outcome function $f_0$ belongs to $\mathcal{F}$ where $\mathcal{F}$ is a closed and convex set of measurable real-valued functions such that for all $f \in \mathcal{F}$, $\mathbb{E}_P[|f|] < \infty, \mathbb{E}_Q[|f|] < \infty$, and $-f \in \mathcal{F}$.
\end{assumption}
For the causal inference problem setting, Assumption 2 requires making an assumption about the relation of the potential outcome $Y(0)$ to the covariates. For the domain adaptation problem setting, the assumption is about the relationship between the accuracy of a predictor, $\ell(h(X),Z)$, to its input features, $X$. 

Many choices of $\mathcal{F}$ in Assumption \ref{outcomes} are quite general and justifiable with domain knowledge. Some examples for $0 < B < \infty$ are:
\begin{align*}
    \text{Bounded functions: } &\mathcal{F}_\infty \coloneqq \{ f : \Vert f \Vert_\infty \leq B \}\\
    \text{Lipschitz functions: } &\mathcal{F}_\text{Lip(c)} \coloneqq
    \{ f : \Vert f \Vert_\text{Lip(c)} \leq B \}\\
    \text{RKHS functions: } &\mathcal{F}_\mathcal{H} \coloneqq \{ f : \Vert f \Vert_\mathcal{H} \leq B \},
\end{align*}
where $\Vert \cdot \Vert_\text{Lip(c)}$ denotes the Lipschitz constant with respect to a metric $c$ and $\Vert \cdot \Vert_\mathcal{H}$ denotes the norm in some Reproducing Kernel Hilbert Space (RKHS), $\mathcal{H}$.

Under Assumption 2, the bias is bounded by the worst-case discrepancy in means over $\mathcal{F}$. This quantity is called an \emph{integral probability metric} (IPM), defined for any set of functions, $\mathcal{G}$, and any $M,N \in \mathcal{M}(\mathcal{X})$ as:\footnote{If $g \in \mathcal{G} \implies -g \in \mathcal{G}$ then the absolute value can be omitted.}
\begin{align*}
    \text{IPM}_{\mathcal{G}}(M, N) \coloneqq \sup_{g \in \mathcal{G}} \big\{ |\mathbb{E}_M [ g ] - \mathbb{E}_N [ g ]| \big\}. 
\end{align*}
The bias term (1) for a re-weighted population $R$ under Assumption 2 is upper-bounded by:
\begin{align}
    \left| \mathbb{E}_Q[ f_0 ] - \mathbb{E}_R[f_0] \right| \leq \text{IPM}_\mathcal{F}(Q,R).\label{biasupperbound}
\end{align}
This value is always finite by our assumptions on $\mathcal{F}$ and we can trade it off against the variance of the weights. 

Before introducing the general form of balancing weights in Section \ref{introbalancing}, we define two quantities that will be useful in our discussion below, the maximum and minimum bias.

\begin{definition*}[Maximum and minimum bias]
The maximum bias, $\delta_{\max}$, is the bias under uniform weights (when $R = P$).
The minimum bias, $\delta_{\min}$, is the smallest bias achieveable by reweighting $P$. 
\begin{align}
    \delta_{\max} &\coloneqq \text{IPM}_\mathcal{F}(Q,P) \label{deltamax} \\[0.5em]
    \delta_{\min} &\coloneqq \inf_{\substack{R \in \mathcal{M}(P) \\ \mathbb{E}_R[1] = 1}} \big\{\text{IPM}_\mathcal{F}(Q,R) \big\}. \label{deltamin} 
\end{align}
\end{definition*}
Since $R = P$ is feasible for (\ref{deltamin}), $\delta_{\min} \leq \delta_{\max}$. In the special case where overlap holds, $R = Q$ is also feasible, which implies $\delta_{\min} = 0$.


\subsection{Minimax Balancing Weights}\label{introbalancing}

Assumption 2 and the resulting IPM bound on the bias (\ref{biasupperbound}) lead to a generalized balancing weights estimator as discussed in \cite{kallus2020generalized} and \cite{eli2021balancingact}. Define $\sigma^2 \coloneqq \sup_{x \in \mathcal{X}} \sigma^2_0(x)$, where we assume $0 < \sigma^2 < \infty$. We can plug these bounds into the MSE to arrive at the following optimization problem:
\begin{align}
    \inf_{\substack{R \in \mathcal{M}(P) \\ \mathbb{E}_R[1] = 1}} \Bigg\{ \text{IPM}_\mathcal{F}&(Q, R)^2 +  \sigma^2\mathbb{E}_P\left[\left(\frac{dR}{dP}\right)^2 \right] \Bigg\}\label{minmse}
\end{align}
A solution always exists because the objective is finite for $R = P$, which is feasible. For $\sigma^2 > 0$, the problem is strongly convex in $R$ and has a unique solution. Since the IPM term is itself a supremum, this estimator is sometimes referred to as minimax balancing weights.

Furthermore, $\exists \delta > 0$ such that (\ref{minmse}) has the same minimizer as: \vspace*{-3mm}
\begin{align}
    \inf_{\substack{R \in \mathcal{M}(P) \\ \mathbb{E}_R[1] = 1}}& \mathbb{E}_P\left[\left(\frac{dR}{dP}\right)^2 \right] \label{constrainedproblem}\\
    \text{ such that }&
    \text{IPM}_\mathcal{F}(Q, R) \leq \delta  .\nonumber
\end{align}
We view $\sigma^2$ and $\delta$ as exchangeable tuning parameters: $\sigma^2$ represents the importance of reducing the variance of the weights; $\delta$ represents the level of acceptable bias. For $\sigma^2 \in (0,\infty)$, the corresponding $\delta$ lies in $(\delta_{\min}, \delta_{\max})$. 

\subsection{Our Contributions}

In this paper, we start from the premise that Assumption 2 is necessary to achieve a reasonable MSE in high dimensions leading to estimators (\ref{minmse}) and (\ref{constrainedproblem}). Our main argument is that Assumption 2 immediately implies two additional results. 

First, we derive a general duality result that lets us rewrite problems (\ref{minmse}) and (\ref{constrainedproblem}) as a single convex optimization problem over $\mathcal{F}$. Therefore, we can solve the minimax balancing weights problem by optimizing a simple convex loss over a function class. Furthermore, this reformulation shows that the optimal weights are always a rescaled and recentered member of $\mathcal{F}$. 

Second, we no longer need an overlap assumption. Before restricting $f_0$, we saw that only $w = dQ/dP$ could guarantee finite bias. Therefore, to bound the MSE we needed the density ratio to exist. But once we assume that $f_0 \in \mathcal{F}$, we no longer need the density ratio to exist, and we can simply minimize our bound on the MSE directly. Moreover, we argue that Assumption 2 provides us with a more appropriate \emph{quantitative} measure of overlap --- the minimum bias, $\delta_\text{min}$ --- that precisely characterizes the difficulty of translating results from one distribution to another.

\section{Duality Theory for Balancing Weights}\label{dualsection}

In this section, we derive a dual characterization of the solution, $R^*$, to problems (\ref{minmse}) and (\ref{constrainedproblem}) and the corresponding minimax weights $w^* = dR^*/dP$.

\subsection{The Variance of the Weights}\label{phiweights}

Our dual derivation uses the fact that the variance term can be written as a special case of a class of information-theoretic divergences called $\phi$-divergences \citep{sriperumbudur2009integral}. 
These have a variational representation that will allow us to simplify the minimax problems (\ref{minmse}) and (\ref{constrainedproblem}) into a single convex loss.

\begin{definition*}[$\phi$-Divergence]
For any convex function $\phi$ with $\phi(1) = 0$, the $\phi$-\emph{divergence} between $M \in \mathcal{M}(\mathcal{X})$ and $N \in \mathcal{P}(\mathcal{X})$ is:
\begin{align*}
    D_\phi(M||N) \coloneqq \mathbb{E}_N \left[ \phi\left(dM/dN\right) \right],
\end{align*}
where $D_\phi(M||N) = \infty$ if $M$ is not absolutely continuous with respect to $N$.
\end{definition*}

Notice that we can subtract off a constant to re-center our variance term in (\ref{constrainedproblem}) without affecting the minimizer over $R$. We can then rewrite the objective as the divergence between $R$ and $P$ with $\phi(x) = x^2 - 1$. This is known as the $\chi^2$ divergence, and we denote it $D_2(R||P)$:
\begin{align*}
    \mathbb{E}_P\left[\left(\frac{dR}{dP}\right)^2 - 1\right] = D_2(R||P).
\end{align*}



\paragraph{Variational representations.}
It is possible to express $\phi$-divergences in a dual form, called a \emph{variational representation}, as a supremum over measurable functions. Let $M \in \mathcal{M}(\mathcal{X})$  and let $N \in \mathcal{P}(\mathcal{X})$. Let $\phi^*$ denote the convex conjugate of $\phi$. Then \cite{keziou2003dual} and \cite{nguyen2005divergences} show that: 
\begin{align}
    D_\phi(M||N) = \sup_{f} \big\{ \mathbb{E}_M[f] - \mathbb{E}_N[\phi^*(f)] \big\},\label{var-nyugen}
\end{align}
where the supremum is over all real-valued measurable functions on $\mathcal{X}$. If we additionally assume, as we do for $R$, that $\mathbb{E}_M[1] = 1$, then we have the tighter representation,
\begin{align}
    D_\phi(M||N) = \sup_{f} \big\{ \mathbb{E}_M[f] - \Lambda_N^\phi[f] \big\}\label{variational}\\
    \text{ where } \Lambda_N^\phi[f] \coloneqq \inf_{\lambda \in \mathbb{R}}\{ \lambda + \mathbb{E}_N[\phi^*(f-\lambda)]\}.\nonumber
\end{align}
This result, using the infimum over $\lambda$ in the spirit of \cite{ruderman2012tighter}, appears to have been independently proposed by \cite{agrawal2020optimal} and \cite{birrell2020optimizing}. Under minimal conditions on $\phi$, the suprema in (\ref{var-nyugen}) and (\ref{variational}) are achieved by $\phi'(dM/dN)$. 

\subsection{Dual Formulation}

We now present our main duality result under Assumption 2 where $f_0 \in \mathcal{F}$. 

\begin{theorem}\label{dualtheorem}
Under Assumptions 1 and 2, for $\delta > \delta_{\min}$, the optimization problem (\ref{constrainedproblem}) has a unique solution,
\begin{align*}
    \frac{dR^*}{dP} = 1 + \left( \frac{ \mathbb{E}_Q[f^*] - \mathbb{E}_P[f^*] - \delta}{\text{Var}_P[f^*]}\right)\left( f^* - \mathbb{E}_P[f^*]\right) ,
\end{align*}
where, for a unique $\mu \geq 0$ corresponding to $\delta$, $f^*$ achieves the following supremum:
\begin{align}
    \sup_{f \in \mathcal{F}}  \Big\{ \mathbb{E}_Q[f]  - \mathbb{E}_P[f] - \frac{\mu}{4} \text{Var}_P[f] \Big\}.\label{dualproblem}
\end{align}
The resulting MSE is:
\begin{align}
    \text{MSE}&(R^*) \leq \delta^2 + \sigma^2 \frac{ (\mathbb{E}_Q[f^*] - \mathbb{E}_P[f^*] - \delta)^2}{\text{Var}_p[f^*]}.\label{resultingmse}  
\end{align}
\end{theorem}


\textbf{Proof Sketch}. The full proof is available in the Appendix. Here we provide a brief outline. In the first step, we show that problem (\ref{constrainedproblem}) is equivalent to:
\[ \sup_{f \in \mathcal{F}}  \Big\{ \mathbb{E}_Q[f] +  \inf_{\substack{R \in \mathcal{M}(P) \\ \mathbb{E}_R[1] = 1}} \{(1/\mu)D_2(R||P) - \mathbb{E}_R[f]\} \Big\} \]
for some $\mu > 0$ corresponding to $\delta$. In the second step, we apply (\ref{variational}) for the $\chi^2$ divergence to show that the inner subproblem has an explicit solution:
\[ \inf_{\substack{R \in \mathcal{M}(P) \\ \mathbb{E}_R[1] = 1}} \{(1/\mu)D_2(R||P) - \mathbb{E}_R[f]\} = - \mathbb{E}_P[f] - \frac{\mu}{4} \text{Var}_P[f]. \] 
The theorem then follows from standard convex duality results. 

\begin{remark}[The Shape of the Weights]
The weights $dR^*/dP$ are equal to $f^*$ multiplied by some scalar $s_1$ and then shifted by some scalar $s_2$:
\begin{align*}
     \frac{dR^*}{dP} = s_1 + s_2 f^*,
\end{align*}
where $s_1$ and $s_2$ depend on both $\delta$ and $f^*$. Therefore, if we assume $\mathcal{F}$ is the set of quadratic functions, then the balancing weights will also be quadratic, and if we assume $\mathcal{F}$ is an RKHS with a certain kernel, then the balancing weights will belong to an RKHS with that same kernel.

\end{remark}

\begin{remark}[Other $\phi$-Divergences]\label{otherphi} 
We can replace the $\chi^2$ divergence in the balancing weight problems (\ref{minmse}) and (\ref{constrainedproblem}) with other $\phi$-divergences. A duality result corresponding to Theorem \ref{dualtheorem} will hold for any convex function $\phi$ such that $\phi(1) = 0$ with convex conjugate $\phi^*$ such that $\{\phi^* < \infty \} = \mathbb{R}$. See the Appendix for details. We can use this general formulation to derive corresponding duality results for entropy balancing \citep{hainmueller2012entropy} or other measures of dispersion \citep[see][]{eli2021balancingact}.
\end{remark}

\begin{remark}[Tuning Parameters] For every $\delta$ there is a unique corresponding $\mu$. Therefore, we can treat $\mu$ as a tuning parameter instead of $\delta$ and solve (\ref{dualproblem}) directly. In terms of $\mu$, the solution to (\ref{constrainedproblem}) is:
\begin{align} \frac{dR^*}{dP} = 1 + \frac{\mu}{2}\left( f^* - \mathbb{E}_P[f^*]\right)\label{muform} \end{align}
and there is a closed form relationship between $\mu$ and $\delta$ given by:
\[ \delta =  \mathbb{E}_Q[f^*] - \mathbb{E}_P[f^*] - \frac{\mu}{2} \text{Var}_P[f^*]. \]
Going forward, we will often use $\delta$ and $\mu$ interchangeably. 
\end{remark}

\subsection{The Full Information Case}

To help illustrate Theorem \ref{dualtheorem}, consider the simplified setting where we know $f_0$ exactly. This corresponds to a special case of Assumption 2 where $\mathcal{F}$ is the convex hull of $\{ f_0, -f_0 \}$. Assume without loss of generality that $\mathbb{E}_Q[f_0] \geq \mathbb{E}_P[f_0]$. Then, applying Theorem \ref{dualtheorem}, we get $f^* = f_0$, and
\begin{align*}
    \frac{dR^*}{dP} = 1 + \left( \frac{ \mathbb{E}_Q[f_0] - \mathbb{E}_P[f_0] - \delta}{\text{Var}_P[f_0]} \right) \left( f_0 - \mathbb{E}_P[f_0]\right).
\end{align*}
The optimal weights are always a rescaled and recentered version of $f_0$. In this special case, the dual optimal $f^*$ does not depend on $\delta$; only the scaling factor does. Therefore, the MSE bound (\ref{resultingmse}) becomes a quadratic in $\delta$ and we can solve for the optimal bias:
\begin{align*}
    \delta^* = \left(\frac{\sigma^2 }{\text{Var}_P[f_0] + \sigma^2} \right) |\mathbb{E}_Q[f_0] - \mathbb{E}_P[f_0]|,
\end{align*}
which gives
\begin{align*}
    \text{MSE}&(w^*) \leq  \left(\frac{\sigma^2}{\text{Var}_P[f_0] + \sigma^2}\right) (\mathbb{E}_Q[f_0] - \mathbb{E}_P[f_0])^2.
\end{align*}
This is an independently interesting result. With complete information, we can analytically find the optimal bias-variance trade-off. Under homoskedasticity, these weights have the smallest possible MSE over all $w$ such that $\mathbb{E}_P[w]= 1$.

\subsection{The Linear Case}

For a second simple example, we return to the linear problem in (\ref{linearbal}). In this case, duality shows that balancing weights are equivalent to fitting a linear model. In fact, for a certain choice of linear $\mathcal{F}$, problem (\ref{dualproblem}) is identical to linear regression. 

Let $g : \mathcal{X} \rightarrow \mathbb{R}^d$ be some feature map. Assume that our balance constraint is: 
\[ \Vert \mathbb{E}_R[g(X)] - \mathbb{E}_Q[g(X)] \Vert_2 \leq \delta. \]
This is equivalent to problem (\ref{constrainedproblem}) using the following linear function class:
\[ f_0 \in \mathcal{F}_\text{lin} = \left\{ \beta^T g(X) : \Vert \beta \Vert_2 \leq 1 \right\}. \]
Applying Theorem \ref{dualtheorem}, we know $f^* = (\beta^*)^T g(X) \in \mathcal{F}_\text{lin}$ and therefore the optimal weights will be linear. Solving (\ref{dualproblem}) via calculus, we get:
\[ \beta^* = c_1 (\text{Cov}_P [ g(X) ] + c_2 I)^{-1} (\mathbb{E}_Q[g(X)] - \mathbb{E}_P[g(X)]) \]
for some scalars $c_1$ and $c_1$ that depend on $\mu$. Notice the dependence on the inverse of the covariance of the features plus a regularization term. This is another way of deriving a well-known result: for $\mathcal{F}_\text{lin}$, problem (\ref{constrainedproblem}) is identical to estimating $\mathbb{E}_Q[Y]$ by fitting a ridge regression in the $P$ population and then applying it to the $Q$ population. See \cite{kallus2020generalized} for a direct proof. 
If we replace the $\ell_2$-norm with the $\ell_1$-norm, we obtain a similar equivalence for Lasso. When the regularization term is $0$, we obtain linear regression as a special case.  

Several other papers have recognized the duality between linear regression and balancing weights estimators for a linear function class; see \cite{zhao2017entropy, zhao2019covariate, wang2020minimal, tan2020regularized, eli2021balancingact}. Theorem \ref{dualtheorem} generalizes these existing duality results for linear function classes to general function classes $\mathcal{F}$.

\section{Outcome Assumptions and Overlap}\label{densityratio}

In this section, we discuss the implications of the outcome assumption for overlap. First, we show that if overlap holds, then under conditions on $\mathcal{F}$, as $\delta \rightarrow 0$, the balancing weights converge to the importance weights $dQ/dP$. 

However, when overlap is violated, the only impact on the balancing weights estimator is that the minimum bias (which depends on our function class $\mathcal{F}$) is greater than zero. Due to Assumption 2, for any $\delta > \delta_{\min}$ there still exists a solution to (\ref{constrainedproblem}) that bounds the MSE. If the variance of the outcomes is large, then we naturally want to choose $\delta$ larger than $\delta_{\min}$ and the failure of overlap has no impact on our estimator. 

Instead, we argue that we should use the minimum bias, $\delta_{\min}$, directly as a measure of practical overlap violations. We illustrate that in finite samples, $\delta_{\min}$ can be large even when overlap holds in the super population, and likewise that $\delta_{\min}$ can be small even when overlap is violated in the super population. Therefore, under Assumption 2, $\delta_{\min}$ is a more precise summary of the underlying difficulty of the reweighting problem.




\subsection{Convergence to Importance Weights}\label{gaussianexample}

\begin{figure}[ht]
    \centering
    \includegraphics[width=0.42\textwidth]{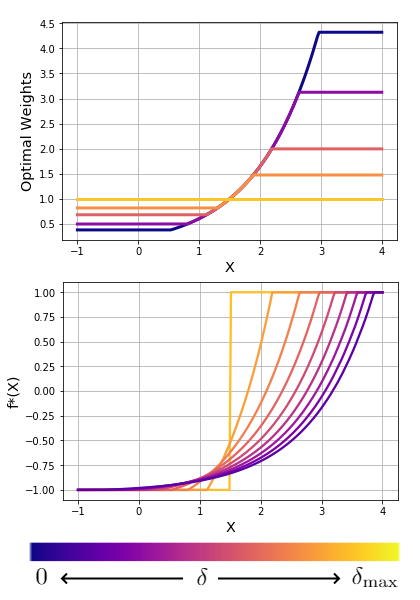}
    \caption{The optimal weights and corresponding dual optimal function for the Gaussian example, with $\delta$ starting at $\delta_{\max}$  and shrinking towards zero.}
    \label{fig:gaussian-weights}
\end{figure}

We begin with an example. Let $\mathcal{X} = \mathbb{R}$. Let $P$ be Gaussian with mean $1$ and variance $1$, let $Q$ be Gaussian with mean $2$ and variance $1$, and let $p$ and $q$ denote their densities. Let $\mathcal{F} = \{f: \Vert f \Vert_\infty \leq 1 \}$ so that the outcome function is bounded between $-1$ and $1$. The solution to the dual problem, $f^*$, and the corresponding optimal weights are illustrated in Figure \ref{fig:gaussian-weights}.

The weights have a distinctive form. When $\delta = \delta_{max}$, the optimal weights are uniform. As the allowed bias $\delta$ decreases, the optimal weights trace out the density ratio $dQ/dP$ but truncated above and below. This is the form of a well-known estimator in the causal inference literature, IPW with a trimmed propensity score \citep{yang2018asymptotic}: under Assumption 2 with bounded functions, the balancing weights formulation provides formal justification for using the truncated density ratio for weights. As $\delta \rightarrow 0$, the optimal weights converge to $dQ/dP$.

In general, convergence to the importance weights will \emph{always} occur as $\delta \rightarrow 0$ under certain conditions on $\mathcal{F}$.

\begin{definition*}[Distribution-defining]
$\mathcal{F}$ is \emph{distribution-defining} if $\forall M,N \in \mathcal{P}(\mathcal{X})$, $\text{IPM}_\mathcal{F}(M,N) = 0$ if and only if $M = N$. 
\end{definition*}
For example, $\mathcal{F}_\infty$ and $\mathcal{F}_\text{Lip(c)}$ are distribution-defining, as is $\mathcal{F}_\mathcal{H}$ for a universal kernel. When $\mathcal{F}$ is distribution-defining then only $dQ/dP$ can achieve worst-case bias zero. Therefore, when overlap holds and $\mathcal{F}$ is distribution-defining, the optimal weights, $w^* \rightarrow dQ/dP$ as $\delta \rightarrow 0$.

This connection between balancing weights and the density ratio is not new: among others, \cite{zhao2019covariate} makes a similar point. We discuss the connections to our duality result in the Appendix.

\subsection{Balancing Weights Without Overlap}

What if overlap does not hold? Then if $\mathcal{F}$ is distribution-defining, by definition, $\delta_{\min} > 0$. In this case, problem (\ref{constrainedproblem}) still has a solution that bounds the MSE for any $\delta \geq \delta_{\min}$, but the failure of overlap precludes us from using $\delta = 0$. However, the motivation behind balancing weights is to avoid using an unbiased estimator: if the variance of the outcomes is sufficiently high, we might still prefer to use $\delta > \delta_{\min}$.

Consider a simple example in which we reweight $P = \text{Uniform}(1,2)$ to target $Q = \text{Uniform}(1.01, 2.01)$. While $Q$ is not absolutely continuous with respect to $P$, intuitively, we should be able to find $w$ that achieves small error because the distributions are close to each other. The function class $\mathcal{F}$ provides a formal definition of ``close to each other'' for the purposes of reweighting. 

For these uniform $P$ and $Q$, there is an irreducible bias, $\delta_{\min}$, for any possible weights:
\[ \delta_{\min} = \sup_{f\in\mathcal{F}} \int_1^{1.01} f(x) dx + \sup_{f\in\mathcal{F}} \int_2^{2.01} f(x) dx \]
If $\mathcal{F}$ is unrestricted, then $f$ could take on arbitrarily large values on the intervals $[1,1.01]$ and $[2,2.01]$. Therefore, the bias is unbounded without Assumption 2, which typically justifies imposing an overlap assumption. However, if we assume $\mathcal{F} \in \{ f : \Vert f \Vert_\infty \leq B \}$, for example, then we have $\delta_{\min} = 0.02 B$ which may be quite small. 

The parameter $\delta > \delta_{\min}$ in problem (\ref{constrainedproblem}) is a tuning parameter that trades off bias and variance. If the variance of the outcomes is very large, then we may prefer to use a value of $\delta$ larger than $\delta_{\min}$. In this case, the overlap violation would not have any impact on our estimator at all. On the other hand, if the variance of the outcomes is small relative to $\delta_{\min}$, we may prefer to use a value of $\delta$ close to $0$. The best we could do would be to set $\delta = \delta_{\min}$; without overlap, the best achievable lower bound for the MSE is $\delta_{\min}^2$.

\subsection{Quantitative Overlap}\label{quantitativeoverlap}

In finite samples, we argue that $\delta_{\min}$ will often be a more useful measure of overlap than the existence of the density ratio in a super-population. For example, let $P_\text{super} = \mathcal{N}(100,1)$ and  $Q_\text{super} = \mathcal{N}(-100,1)$. Technically, overlap holds and the density ratio exists over all of $\mathbb{R}$. For concreteness, let $\mathcal{F} = \mathcal{F}_\mathcal{H}$ be an RKHS with a Gaussian kernel. Then for $P_\text{super}$ and $Q_\text{super}$, $\delta_{\min} = 0$, because $w = dQ/dP$ will perfectly balance the RKHS. However, any finite dataset will have severe practical overlap violations. Let $P$ be a sample of $n$ data points from $P_\text{super}$ and likewise for $Q$. With high probability, the points in $P$ and the points in $Q$ will be far apart, and as a result, $\delta_{\min}$ over the RKHS will be large. 

On the other hand, if we let $P_\text{super} = \text{Uniform}(1,2)$ and $Q_\text{super}= \text{Uniform}(1.01,2.01)$, overlap does not hold and $\delta_{\min}$ will be non-zero for the super-population. But, for corresponding finite samples $P$ and $Q$, $\delta_{\min}$ is still likely to be very small. In these examples, super-population overlap is misleading, whereas $\delta_{\min}$ is a precise quantitative summary of the difficulty of the reweighting problem for function class $\mathcal{F}$.

\section{IHDP Example} \label{ihdp}

In this section, we walk through an example on a real dataset to make the previous two sections more concrete. We apply balancing weights to the Infant Health and Development Program (IHDP) using an RKHS function class.

\subsection{The IHDP Dataset and Setup}

The Infant Health and Development Program (IHDP) data set is a standard observational causal inference benchmark from \citet{hill2011bayesian}, based on data from a randomized control trial of an intensive home visiting and childcare intervention for low birth weight infants born in 1985. 
We consider a non-experimental subset of the original data with $n_0 = 608$ children assigned to control, $n_1 = 139$ children assigned to treatment, and $n = 747$ total children.
For all children, we have a range of baseline covariates, including both categorical covariates, like the mother's educational attainment, and continuous covariates, like the child's birth weight. 
Our goal is to estimate the average outcome (a standardized test score) in the absence of the intensive intervention. We observe this outcome for the 608 control children, and want to re-weight these observations to estimate the missing mean for the 139 treated children.


To do so, we use an RKHS as a flexible but tractable functional form for $f_0$. In particular, we assume that $\mathcal{F} = \mathcal{F}^B_\mathcal{H} \coloneqq \{ f : \Vert f \Vert_\mathcal{H} \leq B \}$ for $B < \infty$, where $\mathcal{H}$ is the RKHS induced by the Gaussian kernel,
\[\mathcal{K}(x_1,x_2) = \exp \left( - \frac{1}{2} \Vert x_1 - x_2 \Vert_2^2 \right).  \]
Define $K \in \mathbb{R}^{n \times n}$ with $K_{ij} = \mathcal{K}(X_i, X_j)$. Then for any $f \in \mathcal{F}$, there exists an $\alpha \in \mathbb{R}^n$ such that $\alpha^TK\alpha \leq B$ and $f(X_j) = \sum_{i=1}^n \alpha_i K_{ij}, \forall j.$

\subsection{Solving the Dual Problem}

We compute the minimax balancing weights by solving the dual problem (\ref{dualproblem}) directly for many values of the tuning parameter $\mu > 0$.  The dual problem can be written as a quadratic optimization problem over the vectors $\alpha$ that characterize the $f \in \mathcal{F}$. See the Appendix for details. We obtain the corresponding optimal weights by plugging the resulting $f^*$ into (\ref{muform}).

\begin{figure}[ht]
    \centering
\includegraphics[width=0.47\textwidth]{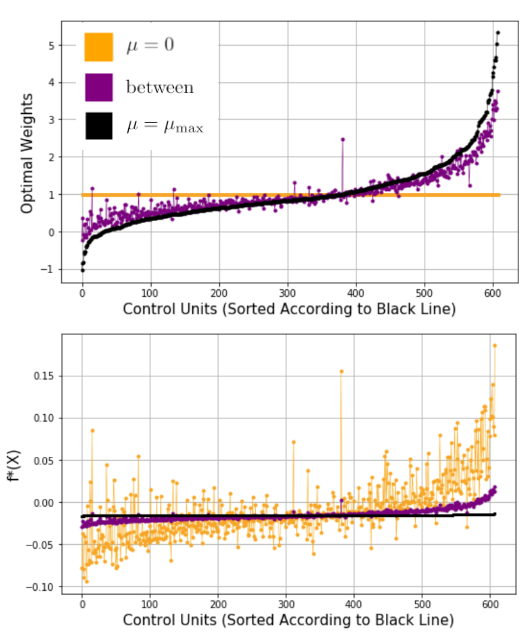}
    \caption{The optimal weights and corresponding dual optimal function for the IHDP example for the extreme values of $\mu$ and one intermediate value.}
    \label{fig:ihdp-mmd}
\end{figure}

The balancing weights interpolate between two extremes. See Figure \ref{fig:ihdp-mmd} for an illustration. At one extreme are the weights with maximum bias and minimum variance. This is achieved at $\mu = 0$, which results in uniform weights and corresponding bias $\delta = \delta_{\max}$.

At the other extreme are the weights with maximum variance and minimum bias. Since some of the covariates are continuous, the data points for the control and treated groups have disjoint support. Therefore, there are no weights that achieve zero bias. Instead, we find weights that achieve the smallest possible bias over $\mathcal{F}^B_\mathcal{H}$, $\delta_{\min}$, which will correspond to some  $\mu = \mu_{\max} < \infty$. We find $\mu_{\max}$ by increasing $\mu$ until the bias stops decreasing. The corresponding weights are shown in black in Figure \ref{fig:ihdp-mmd}.

The function class $\mathcal{F}^B_\mathcal{H}$ has the nice property that the worst-case bias scales with the norm bound $B$, $\text{IPM}_{\mathcal{F}^B_\mathcal{H}}(R,Q) = B \cdot \text{IPM}_{\mathcal{F}^1_\mathcal{H}}(R,Q)$. Furthermore, regardless of $B$, the optimal weights remain identical. Therefore, we can report the bias as a fraction of the size of functions in $\mathcal{F}$. For the IHDP data, $\delta_{\max} = 0.102 B$ and $\delta_{\min} = 0.089 B$ with corresponding variances $\sigma^2$ (with uniform weights) and $1.7 \sigma^2$. For this particular problem, we achieve most of the bias reduction with smaller weights: the intermediate weights in Figure \ref{fig:ihdp-mmd} have $\delta = 0.090 B$ with variance $1.35 \sigma^2$, highlighting the relevance of the bias-variance trade-off. 

In any real data set with continuous covariates, two finite samples will typically have disjoint support like we have here. The standard approach in causal inference is to assume that overlap holds in the super-populations from which the samples were drawn. In this case, we could approximate the density ratio asymptotically. However, as we emphasized in Section \ref{quantitativeoverlap}, the implications of overlap for balancing weights are entirely summarized by $\delta_{\min}$ so we do not need to make such an assumption. 


\begin{remark}[Computational Advantages of the Dual]
For an RKHS, there is a closed form of the IPM, which makes the primal and dual problems equally easy to solve. But in some situations, it is computationally easier to solve the dual problem (\ref{dualproblem}) directly instead of the primal problem (\ref{constrainedproblem}). Consider a class of neural networks parameterized by bounded network weights $\theta$. Then handling the IPM constraint in the primal problem requires adversarial training, as in \cite{kallus2020deepmatch}, which can be quite computationally challenging. On the other hand, (\ref{dualproblem}) requires training a neural network once with a convex loss function which can be accomplished with off-the-shelf SGD.  
\end{remark}

\section{Robustness}


Balancing weights rely heavily on the function class in Assumption 2. In this section, we show that with minimal moment conditions we can still retain a bound on the bias even if we have misspecified the function class $\mathcal{F}$. We consider two functions classes. First, a \emph{misspecified} $\mathcal{F}$ for which we solve (\ref{constrainedproblem}) to find $R^*$ such that $\text{IPM}_\mathcal{F}(Q,R^*) \leq \delta$. Second, the \emph{true} function class, $\mathcal{G}$ such that $f_0 \in \mathcal{G}$ and $f_0 \notin \mathcal{F}$. To bound the bias, we need to show that
\begin{align} \text{IPM}_\mathcal{F}(Q,R^*) \leq \delta \implies \text{IPM}_\mathcal{G}(Q,R^*) \leq \rho(\delta)\label{robustcondition} \end{align}
for some $\rho < \infty$ which has good scaling with $\delta$. Without further assumptions, (\ref{robustcondition}) will \emph{not} hold for any $\mathcal{G}$.

IPMs correspond to common perturbations in the robust statistics literature. For example, $\text{IPM}_{\mathcal{F}_\infty}$ and $\text{IPM}_{\mathcal{F}_\text{Lip(c)}}$ are equivalent to the total variation (TV) distance and Wasserstein distance respectively. For $\mathcal{F} = \mathcal{F}_\infty$, we can apply Lemma E.2 from \cite{zhu2019generalized} to achieve (\ref{robustcondition}) for any $\mathcal{G}$. We require an Orlicz norm bound under $Q$ and $R^*$ on $g(X)$ for all $g \in \mathcal{G}$. For a simple example, let $\mathcal{G}$ be linear. Then we get the following result:

\begin{prop}
Let $TV(Q, R^*) \leq \delta$ and let $f_0 \in \{\beta^Tx : \Vert \beta \Vert \leq 1 \}.$ If $R^*$ and $Q$ have bounded covariance, then we have the following upper bound on the bias:
\[ | \mathbb{E}_Q[f_0] - \mathbb{E}_{R^*}[ f_0 ] | \leq \rho_1(\delta), \]
where $\rho_1(\delta) = O(\sqrt{\delta})$.

If instead $R^*$ and $Q$ are sub-Gaussian, then we have the following upper bound on the bias:
 \[ | \mathbb{E}_Q[f_0] - \mathbb{E}_{R^*}[ f_0 ] | \leq \rho_2(\delta), \]
where $\rho_2(\delta) = O(\delta\sqrt{\log(1/\delta)})$.
\end{prop}


For general $\mathcal{G}$, the rate of $\rho$ in terms of $\delta$ is similar, but the moment conditions on $X$ become stronger. In practice, these robust statistics results mean that we can make a best guess about $\mathcal{F}$ and as long as $Q$ is sufficiently ``nice'', the true bias will not be much larger than $\delta$.

\bibliographystyle{abbrvnat}
\bibliography{sample_paper}

\clearpage
\appendix

\thispagestyle{empty}

\onecolumn \makesupplementtitle




\textbf{CONTENTS}

\begin{enumerate}
    \item[A] Proof of Theorem \ref{dualtheorem}
    \item[B] General statement and proof for $\phi$-divergences for Remark \ref{otherphi}
    \item[C] Extension with non-negative weights
    \item[D] Connection to surrogate loss for density ratio estimation
    \item[E] Details of RKHS optimization problem in Section \ref{ihdp}
\end{enumerate}

\vfill
\newpage

\section{Proof of Theorem \ref{dualtheorem}}

We derive a dual formulation of the optimization problem:
\begin{align*}
    \inf_{\substack{R \in \mathcal{M}(P) \\ \mathbb{E}_R[1] = 1}}& D_2(R||P) \\
    \text{ such that }&
    \text{IPM}_\mathcal{F}(Q, R) \leq \delta ,
\end{align*}
where $\delta > \delta_{\min}$.  As a reminder, $D_2(R||P) \coloneqq \mathbb{E}_P[(dR/dP)^2 - 1]$ is the $\chi^2$ divergence  and $\text{IPM}_\mathcal{F}(Q,R) \coloneqq \sup_{f \in \mathcal{F}}\{ \mathbb{E}_Q[f] - \mathbb{E}_R[f] \}$. Note that this problem takes the form of a projection in $D_2$ of $P$ onto an IPM ball around $Q$.

By the definition of $\delta_{\min}$, the constraint set is non-empty and convex. $D_2$ is strictly convex in $R$ and $0 \leq D_2(R||P) < \infty$ so there is a unique solution. 

When $P$ already satisfies the IPM constraint, then $R=P$ has objective $0$ and we're done (i.e. we don't need to do a projection, $P$ is already on or in the ball). Otherwise, by standard use of the Lagrangian, we claim (details in Section \ref{lagrange} below) that for some $\mu > 0$ corresponding to $\delta$, this problem is equivalent to:
\begin{align*}
    \inf_{\substack{R \in \mathcal{M}(P) \\ \mathbb{E}_R[1] = 1}} \big\{ (1/\mu) D_2(R||P) + \sup_{f \in \mathcal{F}} \{ \mathbb{E}_Q[f] - \mathbb{E}_R[f] \}  \big\}.
\end{align*}

\subsection{Exchanging hard subproblem for an easy subproblem}

The inner supremum is hard to solve for arbitrary $\mathcal{F}$. However, we can make a series of transformations to get an easier subproblem with a closed-form solution:

\begin{align*}
     &\inf_{\substack{R \in \mathcal{M}(P) \\ \mathbb{E}_R[1] = 1}} \big\{ (1/\mu) D_2(R||P) + \sup_{f \in \mathcal{F}} \{ \mathbb{E}_Q[f] - \mathbb{E}_R[f] \}  \big\}\\
      &= \inf_{\substack{R \in \mathcal{M}(P) \\ \mathbb{E}_R[1] = 1}} \sup_{f \in \mathcal{F}} \Big\{ (1/\mu)D_2(R||P) +  \mathbb{E}_Q[f] - \mathbb{E}_R[f] \Big\}\\
      &= \sup_{f \in \mathcal{F}} \inf_{\substack{R \in \mathcal{M}(P) \\ \mathbb{E}_R[1] = 1}}  \Big\{ (1/\mu)D_2(R||P) +  \mathbb{E}_Q[f] - \mathbb{E}_R[f] \Big\}\\
      &= \sup_{f \in \mathcal{F}}  \Big\{ \mathbb{E}_Q[f] +  \inf_{\substack{R \in \mathcal{M}(P) \\ \mathbb{E}_R[1] = 1}} \{(1/\mu)D_2(R||P) - \mathbb{E}_R[f]\} \Big\}
\end{align*}

The only non-trivial step is the interchange of the $\inf$ and the $\sup$. This follows by Sion's Minimax Theorem \citep{sion1958general}. We assumed that $\mathcal{X}$ was a separable Banach space so we have the necessary topological properties. The objective on the second line is continuous and strictly convex in $R$ and is linear in $f$. The set $\mathcal{F}$ is convex and closed and $R$ is in a linear subspace. Furthermore, we know there is a unique solution $R^*$, and so we can always find the necessary compact subset of the linear subspace for $R$ to apply the theorem  \citep[e.g. ala][]{ha1981noncompact}.

\subsection{Solving the easy subproblem with the variational representation}

Next, we apply the variational representation of $\phi$-divergences to get a dual formulation of the inner sub-problem over $R$. Define: $\phi(x) = (1/\mu)(x^2-1)$ which has convex conjugate $\phi^*(y) = (\mu/4)y^2 + (1/\mu)$. The Lagrangian for the infimum for a fixed $f$ is:
\begin{align*}
    \mathcal{L}_f(R,\lambda) = D_\phi(R||P) - \mathbb{E}_R[f - \lambda] - \lambda
\end{align*}

We get the first-order condition:
\begin{align*}
    \phi'(dR^*/dP) = f - \lambda^* \implies \frac{dR^*}{dP} = \frac{\mu}{2}(f - \lambda^*)
\end{align*}

where $\lambda^*$ solves the supremum $\sup_{\lambda \geq 0} g(\lambda)$ over the dual function:
\begin{align*}
     g(\lambda) &\coloneqq - \lambda + \inf_{R \in \mathcal{M}(P)} \{ D_\phi(R||P) - \mathbb{E}_R[f-\lambda] \}\\
     &= - \lambda - \sup_{R \in \mathcal{M}(P)} \{  \mathbb{E}_R[f-\lambda] - D_\phi(R||P) \}\\
     &= - \lambda - D_\phi^*(f - \lambda),
\end{align*}
where $D^*_\phi$ is the convex conjugate of the $\phi$-divergence as a function of $R$ for a fixed $P$. We can then use the standard result \citep[Proposition 4.2][]{broniatowski2006minimization}, $D^*_\phi(f) = \mathbb{E}_P[\phi^*(f)]$. 

Using this form of the dual function, we can write our subproblem over $R$ as:
\begin{align*}
    \inf_{\substack{R \in \mathcal{M}(P) \\ \mathbb{E}_R[1] = 1}} \{(1/\mu)D_2(R||P) - \mathbb{E}_R[f]\} = \sup_{\lambda \geq 0} \{ - \lambda - \mathbb{E}_P[\phi^*(f - \lambda)] \}
\end{align*}

Plugging in $\phi^*$, we can solve for $\lambda^*$ by straightforward calculus:
\begin{align*}
    &\lambda^* = \mathbb{E}_P[f] - \frac{2}{\mu}.
\end{align*}
Now using the first-order conditions, we can find the optimal $R^*$:
\begin{align*}
    \frac{dR^*}{dP} = \frac{\mu}{2}(f - \mathbb{E}_P[f]) + 1
\end{align*}
and after some algebra, a closed form of the subproblem:
\begin{align*}
    \inf_{\substack{R \in \mathcal{M}(P) \\ \mathbb{E}_R[1] = 1}} \{(1/\mu)D_2(R||P) - \mathbb{E}_R[f]\} = - \mathbb{E}_P[f] - \frac{\mu}{4} \text{Var}_P[f].
\end{align*}

\subsection{Writing the original problem as a single optimization problem over $\mathcal{F}$}

Finally, we substitute this form of the sub-problem into our original optimization problem:
\begin{align*}
    &\inf_{\substack{R \in \mathcal{M}(P) \\ \mathbb{E}_R[1] = 1}} \big\{ (1/\mu) D_2(R||P) + \sup_{f \in \mathcal{F}} \{ \mathbb{E}_Q[f] - \mathbb{E}_R[f] \}  \big\}\\
    &= \sup_{f \in \mathcal{F}}  \Big\{ \mathbb{E}_Q[f] +  \inf_{\substack{R \in \mathcal{M}(P) \\ \mathbb{E}_R[1] = 1}} \{(1/\mu)D_2(R||P) - \mathbb{E}_R[f]\} \Big\}\\
    &= \sup_{f \in \mathcal{F}}  \Big\{ \mathbb{E}_Q[f] - \mathbb{E}_P[f] - \frac{\mu}{4} \text{Var}_P[f] \Big\}
\end{align*}
and therefore by duality:
\begin{align*}
    \frac{dR^*}{dP} = \frac{\mu}{2}(f^* - \mathbb{E}_P[f^*]) + 1
\end{align*}
where $f^*$ achieves this supremum.

\subsection{Recovering $\delta$ in terms of $\mu$}

Most of the proof of the theorem is complete. We just need to rewrite $\mu$ in terms of the original tuning parameter $\delta$. Remember from the projection perspective, that $\mu > 0$ corresponds to $P$ outside of the IPM ball. As a result:
\begin{align*}
    \delta = \sup_{f \in \mathcal{F}} \{ \mathbb{E}_Q[f] - \mathbb{E}_{R^*}[f] \} 
\end{align*}

We just proved that $R^*$ achieves the infimum of the objective which equals the supremum of the dual:
\begin{align*}
    &  \text{IPM}_\mathcal{F}(Q,R^*) + (1/\mu) D_2(R^*||P)  \\
    &=  \mathbb{E}_Q[f^*] - \mathbb{E}_P[f^*] - \frac{\mu}{4} \text{Var}_P[f^*]\\
    &= \Big( \mathbb{E}_Q[f^*] - \mathbb{E}_{R^*}[f^*] \Big) + \Big( \mathbb{E}_{R^*}[f^*] - \mathbb{E}_P[f^*] - \frac{\mu}{4} \text{Var}_P[f^*] \Big)
\end{align*}

But then, using the variational representation of the $\phi$-divergence and the definition of the IPM we have:
\begin{align*}
    & \text{IPM}_\mathcal{F}(Q,R^*) + (1/\mu) D_2(R^*||P) \\
    &= \sup_{f \in \mathcal{F}} \Big\{ \mathbb{E}_Q[f] - \mathbb{E}_{R^*}[f] \Big\} + \sup_f \Big\{ \mathbb{E}_{R^*}[f] - \mathbb{E}_P[f] - \frac{\mu}{4} \text{Var}_P[f^*] \Big\}\\
    &= \Big( \mathbb{E}_Q[f^*] - \mathbb{E}_{R^*}[f^*] \Big) + \Big( \mathbb{E}_{R^*}[f^*] - \mathbb{E}_P[f^*] - \frac{\mu}{4} \text{Var}_P[f^*] \Big)
\end{align*}
which implies
\[  \delta = \sup_{f \in \mathcal{F}} \Big\{ \mathbb{E}_Q[f] - \mathbb{E}_{R^*}[f] \Big\} = \mathbb{E}_Q[f^*] - \mathbb{E}_{R^*}[f^*]  \]
Finally, substituting the form of $R^*$ in terms of $f^*$ we get
\begin{align*}
    \delta &= \mathbb{E}_Q[f^*] - \mathbb{E}_P[f^*] - \frac{\mu}{2}\text{Var}_P[f^*]\\
    \implies \mu &= 2 \left(\frac{\mathbb{E}_Q[f^*] - \mathbb{E}_P[f^*] - \delta}{\text{Var}_P[f^*]}\right),
\end{align*}
which concludes the proof.




\subsection{Transformation from $\delta$ to $\mu$ via Lagrangian}\label{lagrange}

Here we provide the details for our earlier claim that we can rewrite the problem over $\delta$ as a problem over $\mu$. The dual function corresponding to the original $\delta$ problem is:
\begin{align*} g(\mu) &=  \inf_{\substack{R \in \mathcal{M}(P) \\ \mathbb{E}_R[1] = 1}} \mathcal{L}(R, \mu)\\
&= \inf_{\substack{R \in \mathcal{M}(P) \\ \mathbb{E}_R[1] = 1}} \{ D_2(R||P) + \mu (\sup_{f \in \mathcal{F}} \{ \mathbb{E}_Q[f] - \mathbb{E}_R[f] \} - \delta) \}
\end{align*}
where $\mathcal{L}$ is the Lagrangian. Notice that the original optimization problem has a strictly convex objective. Furthermore, since the function class $\mathcal{F}$ is convex and closed, $\delta > \delta_{\min}$, and the individual constraints for $f \in \mathcal{F}$ are all linear, the feasible set is convex with a non-empty interior. Then by standard convex duality there exists $R^*$ and $\mu^* \geq 0$ such that $R^*$ solves the original optimization problem, $\mu^*$ achieves $\sup_{\mu \geq 0} g(\mu)$, and $R^*$ achieves the infimum inside $g(\mu^*)$. 

By complementary slackness, $\mu^* = 0$ only when the worst-bias constraint doesn't bind which only occurs when $R=P$ already satisfies the IPM constraint. Then $R=P$ has minimum variance and we're done. So we only need to consider the case where $\mu^* > 0$. 

At $\mu = \mu^*$, the solution to:
\begin{align*}
    \inf_{\substack{R \in \mathcal{M}(P) \\ \mathbb{E}_R[1] = 1}} \big\{ D_2(R||P) + \mu \sup_{f \in \mathcal{F}} \{ \mathbb{E}_Q[f] - \mathbb{E}_R[f] \}  \big\}
\end{align*}
has the same solution as the original problem. Furthermore, since $\mu^* > 0$, we can apply one more  transformation without affecting the infimum to get:
\begin{align*}
    \inf_{\substack{R \in \mathcal{M}(P) \\ \mathbb{E}_R[1] = 1}} \big\{ (1/\mu) D_2(R||P) + \sup_{f \in \mathcal{F}} \{ \mathbb{E}_Q[f] - \mathbb{E}_R[f] \}  \big\}
\end{align*}









\vfill
\newpage
\section{General Statement and Proof for Remark 3.1}

\begin{theorem*}[Birrell et al]
Let $\phi$ be a convex function such that $\phi(1) = 0$ with convex conjugate $\phi^*$ such that $\{ \phi^* < \infty \} = \mathbb{R}$. Let $\phi_\mu$ denote the weighted function $\phi_\mu(x) = (1/\mu)\phi(x)$ and $\phi_\mu^*$ its convex conjugate. Then under Assumption 2, for $\delta > 0$, $\exists \mu \geq 0$ such that the optimization problem,
\begin{align*}
    \inf_{\substack{R \in \mathcal{M}(P) \\ \mathbb{E}_R[1] = 1}}& D_\phi(R||P) \\
    \text{ such that }&
    \text{IPM}_\mathcal{F}(Q,R) \leq \delta, \nonumber
\end{align*}
has a solution,
\begin{align*}
    &R^* = P \text{ when $\mu = 0$, }\\
    &\frac{dR^*}{dP} = (\phi_\mu^*)'(f^* - \lambda^*) \text{ otherwise,}
\end{align*}
where $f^*$ and $\lambda^*$ achieve the supremum,
\begin{align*}
    \sup_{f \in \mathcal{F}} \left\{ \mathbb{E}_Q[f] - \inf_{\lambda \in \mathbb{R}}\{ \lambda + \mathbb{E}_P[\phi_\mu^*(f-\lambda)]\}\right\}.
\end{align*}
\end{theorem*}

The proof begins with an argument identical to 1.5 above which gives the equivalent optimization problem:
\begin{align*}
    \inf_{\substack{R \in \mathcal{M}(P) \\ \mathbb{E}_R[1] = 1}} \big\{  D_{\phi_\mu}(R||P) + \text{IPM}_\mathcal{F}(Q,R)  \big\}.
\end{align*}
From here, conceptually, the proof is similar to above, except we cannot apply our proof directly because general $\phi$ loses some of the nice properties of the quadratic. Instead, using the theory of infimal convolutions, \cite{birrell2020f} prove that for Polish $\mathcal{X}$ and any $\phi$-divergence such that $\{ \phi^* < \infty \} = \mathbb{R}$:
\begin{align*}
    &\inf_{\substack{R \in \mathcal{M}(P) \\ \mathbb{E}_R[1] = 1}} \big\{  D_{\phi}(R||P) + \text{IPM}_\mathcal{F}(Q,R)  \big\}\\
    &= \sup_{f \in \mathcal{F}} \left\{ \mathbb{E}_Q[f] - \inf_{\lambda \in \mathbb{R}}\{ \lambda + \mathbb{E}_P[\phi^*(f-\lambda)]\}\right\}
\end{align*}
and so in particular, it holds for $\phi_\mu$ above.

This result follows from Theorems 2.15 and 3.3 in \cite{birrell2020f} with one minor modification: we do not require that $\lim_{y \rightarrow -\infty} \phi^*(y) < \infty$ which results in $R^*$ no longer being a probability measure because we lose statement (174) in their proof of Theorem C.6. This is in the spirit of the arguments in \citep{broniatowski2006minimization}. In fact, since the original problem is a projection of $P$ in $\phi$-divergence onto an IPM, we can interpret this as a version of \citep{broniatowski2006minimization} Theorem 5.1 which applies for the case of finitely-many linear inequality constraints, generalized to the case of a linear inequality constraint for each $f$ in a convex and closed set $\mathcal{F}$.

\vfill
\newpage

\section{Extension with non-negative weights}

We can use the general Theorem to immediately get results for the case where we require non-negative weights. This is identical to taking $\phi(x) = x^2 - 1$, but restricting the domain to $[0,\infty)$. Then we have
\[ \phi^*_\mu(y) = \frac{\mu}{4} y^2 \mathbf{1}(y \geq 0) + \frac{1}{\mu}  \]
and applying the theorem, we get the dual formulation:
\begin{align*}
    \frac{dR^*}{dP} = \frac{\mu}{2} (f^* - \lambda^*) \mathbf{1}(f^* \geq \lambda^*)
\end{align*}
such that:
\begin{align*}
    \mathbb{E}_P\left[\frac{\mu}{2} (f^* - \lambda^*) \mathbf{1}(f^* \geq \lambda^*)\right] = 1
\end{align*}
and $f^*$ and $\lambda^*$ solve:
\begin{align*}
    \sup_{f \in \mathcal{F}} \left\{ \mathbb{E}_Q[f] - \inf_{\lambda \in \mathbb{R}}\left\{ \lambda + \frac{\mu}{4} \mathbb{E}_P\left[ (f-\lambda)^2 \mathbf{1}(f \geq \lambda) \right] \right\} + \frac{1}{\mu}\right\}.
\end{align*}
The minimax weights are extremely simliar to the minimax weights from Theorem 3.1. In Theorem 3.1 we found a function $f^*$, such that we de-meaned it, rescaled it, and then shifted the result to get weights with expectation equal to 1. In the case where the weights are non-negative, we can no longer just de-mean and add 1. Instead, we have to optimize over all shifts $\lambda$ which give expectation 1 \emph{after} truncating at $0$.

\vfill
\newpage

\section{Connection to surrogate loss for density ratio estimation}

A closely related literature implements regularized estimators of the density ratio via a surrogate loss as first proposed in \cite{nguyen2010estimating}. Similar connections have been made for balancing weights; see the relation to propensity score estimation in \cite{zhao2019covariate, eli2021balancingact}. Here, we compare our dual formulation based on variational representations to density ratio estimation with a surrogate loss.  

Consider the variational representation (\ref{variational}). Specializing to $\phi(x) = \mu(x^2-1)$ for $\mu > 0$, we can write the weighted $\chi^2$ divergence between $Q$ and $P$ as:
\begin{align*}
    \frac{1}{\mu}D_2(Q||P) = \sup_{f}  \Big\{ \mathbb{E}_Q[f]  - \mathbb{E}_P[f] - \frac{\mu}{4} \text{Var}_P[f] \Big\},
\end{align*}
where the supremum is over all real-valued measurable functions. If overlap holds then the supremum is achieved by $2\mu(dQ/dP)$, otherwise $D_2(Q||P) = \infty$. The variational representation is identical to (\ref{dualproblem}), except that the problem in (\ref{dualproblem}) is restricted to functions in $\mathcal{F}$. We immediately have the following corollary:
\begin{corollary}
Under the conditions in Theorem \ref{dualtheorem},
\begin{align*}
    \frac{2}{\mu} \frac{dQ}{dP} \in \mathcal{F} \implies f^* = \frac{2}{\mu} \frac{dQ}{dP} \text{ and } \frac{dR^*}{dP} = \frac{dQ}{dP}.
\end{align*} 
\end{corollary}
If the density ratio exists and a scaled version belongs to the outcome function class, then it is minimax optimal to reweight so that there is zero bias. If $\mathcal{F}$ is sufficiently flexible, then this will always hold as $\mu \rightarrow \infty$, as in the example from Section \ref{gaussianexample}.

There is a clear similarity to density ratio estimation using $\phi$-divergences as in \citet{nguyen2010estimating}. The form of their final estimator looks the same as ours: maximizing a variational representation over a function class. However, our motivations and assumptions are entirely different. \cite{nguyen2010estimating} assumes that the density ratio exists and that it belongs to $\mathcal{F}$. We do not make any assumption about the density ratio. It could have some arbitrary functional form. We show that the shape of the optimal weights is determined by the shape of $f_0$, \emph{not} by the shape of $dQ/dP$. 

If the density ratio exists, and our dual problem is solved over all measurable functions, the only unique solution will be $dQ/dP$. In that sense, once we restrict to optimization over $\mathcal{F}$, it might be helpful to think of the balancing weights as a projection of the density ratio onto our outcome function class. However, in general, we do not need a functional form for the density ratio to specified. In fact, we do not need even require that the density ratio exists. 

\vfill
\newpage

\section{RKHS optimization problem}

The optimization problem that implements our dual formulation using an RKHS for the IHDP dataset is:

\begin{align*}
    &\sup_{f \in \mathcal{F}_\mathcal{H}^B}  \Big\{ \mathbb{E}_Q[f]  - \mathbb{E}_P[f] - \frac{\mu}{4} \text{Var}_P[f] \Big\}\\
    &= \sup_{ \substack{f =  K \alpha \\ \alpha \in \mathbb{R}^n : \alpha^T K \alpha \leq B} } \Big\{ f^T e_q - f^T e_p - ( f^T I_p f - (f^T e_p)^2 )\Big\}
\end{align*}
where:
\begin{itemize}
    \item $e_p \in \mathbb{R}^n$ is a vector equal to $1/n_0$ for those indices corresponding to control group data points and $0$ otherwise
    \item $e_q \in \mathbb{R}^n$ is a vector equal to $1/n_0$ for those indices corresponding to treatment group data points and $0$ otherwise
    \item $I_p \in \mathbb{R}^{n \times n}$ is the matrix with $e_p$ on the diagonal.
\end{itemize}

We solve this problem using the scipy Python package.

\end{document}